\documentclass{PoS}
\usepackage{amssymb,amsmath,graphicx,xcolor,cancel,slashed}

\newcommand{\CPV}{$\cancel{\text{CP}}$}

\title{Present and future prospects for lattice QCD calculations of matrix elements for nEDM}

\ShortTitle{CP Violation and the neutron EDM}

\author{\speaker{Rajan Gupta}\thanks{Work done as part of the 
    PNDME Collaboration whose other members are Tanmoy Bhattacharya, Vincenzo
    Cirigliano and Boram Yoon}
  \\ Theoretical Division T-2, Los Alamos National Laboratory, Los
  Alamos, NM 87545, USA\\ E-mail: \email{rg@lanl.gov}}

\abstract{A status report on the calculations of the contribution of
  four CP violating operators, the $\Theta$-term, the quark EDM, the
  chromo EDM and the Weinberg operator to the neutron EDM are
  presented. At this time, there exit precise physical results only
  for the quark EDM operator by the PNDME collaboration. First results
  showing signal in the contributions of the $\Theta$-term and the
  connected part of the chromo EDM operator have been presented. The
  challenge of divergent mixing in the chromo EDM and Weinberg
  operators has motivated calculations in the gradient flow
  scheme. While there has been steady progress, the challenges
  remaining are large. Results with $O(50\%)$ uncertainty with control
  over all systematic errors can be expected for the $\Theta$-term
  over the next five years. Prediction of a timeline for progress on
  the chromo EDM and the Weinberg operators will depend on when the
  renormalization and divergent mixing of these with lower
  dimensional operators is brought under control. The most optimistic
  scenario is that the gradient flow scheme provides a solution to the
  numerical signal and mixing problems for both the gluonic and quark
  operators.  }

\FullConference{23rd International Spin Physics Symposium - SPIN2018 -\\
		10-14 September, 2018\\
		Ferrara, Italy}

\begin{document}

\section{Introduction}
\label{sec:intro}

One of the deepest mysteries of the observed universe is the
matter-antimatter asymmetry.  The observed universe has
\(6.1^{+0.3}_{-0.2}\times 10^{-10}\) baryons for every black body
photon~\cite{Bennett:2003bz}, whereas in a baryon symmetric universe,
we expect no more that about \(10^{-20}\) baryons for every
photon~\cite{Kolb:1990vq}.  It is difficult to include such a large
excess of baryons as an initial condition in an inflationary
cosmological scenario~\cite{Coppi:2004za}.  The way out of the impasse
lies in generating the baryon excess dynamically (baryogenesis, leptogenesis, ...) during
the evolution of the universe.

In the early history of the universe, if the matter-antimatter
symmetry was broken post inflation and reheating, then one is faced
with Sakharov's three necessary conditions~\cite{Sakharov:1967dj}: the
process has to violate baryon number, evolution has to occur out of
equilibrium, and CP (or equivalently time reversal invariance if CPT
remains unbroken) has to be violated.

\CPV\  exists in the electroweak sector of the standard model (SM) of
particle interactions due to a phase in the Cabibbo-Kobayashi-Maskawa
(CKM) quark mixing matrix~\cite{Kobayashi:1973fv}, and possibly by a
similar phase in the leptonic sector, given that the neutrinos are not
massless~\cite{Nunokawa:2007qh}. The strength of the \CPV\  in the CKM
matrix is much too small to explain baryogenesis. Leptogenesis from
the neutrino sector with large lepton-baryon conversion is another
possible mechanism, however, no CP violation in the lepton sector has
been observed so far.

The SM has an additional source of CP violation arising from the
effect of QCD instantons.  The presence of these finite action
non-perturbative configurations in a non-Abelian theory often leads to
inequivalent quantum theories defined over various
`$\Theta$'-vacua~\cite{Dolgov:1991fr}. However, because of asymptotic
freedom, all non-perturbative configurations including instantons are
strongly suppressed at high temperatures where rates of baryon number
violating processes are sizable. Because of this, \CPV\  due to such
vacuum effects do not lead to appreciable baryon number production. \looseness-1

In short, the overriding consensus is that additional much larger
\CPV\ is needed from physics beyond the SM (BSM).  Even though the BSM
theory that describes nature above the TeV scale is not known, using
the tools of effective field theory one can organize, by symmetry and
dimension, possible \CPV\ interactions at the hadronic scale. One then
needs to quantify their contribution to the neutron electric dipole
moment (nEDM). Each contribution consists of a product of the coupling
(BSM model dependent) and the matrix element of the low-energy
effective interaction (both defined at the hadronic scale0 within the
neutron state (BSM model independent).  In this review, I will discuss the
status of lattice QCD calculation of the matrix elements of four of
the leading, within the effective field theory framework,
\CPV\ operators.

%%%%%%%%%%%%%%%%%%%%%%%%%%%%%%%%%%%%%%%%%%%%%%%%%%%%%%%%%%%%%%%%%%%%%%%%%%%%%%%%
\section{Nucleon Matrix elements of  \CPV\  operators}
\label{sec:SSM}
%%%%%%%%%%%%%%%%%%%%%%%%%%%%%%%%%%%%%%%%%%%%%%%%%%%%%%%%%%%%%%%%%%%%%%%%%%%%%%%%

Over the past few decades, many extensions of the SM have been
proposed in the literature.  At the hadronic scale (${}\sim2$~GeV),
the effects of BSM scenarios that involve heavy degrees of freedom at
the mass scale $\Lambda_{\rm BSM} > M_W$ can be described in terms of effective
local operators composed of quarks and gluons.  Using tools of
effective field theory, one can organize all possible effective \CPV\ 
interactions of quarks and gluons based on symmetry and dimension, and
independent of specific BSM theory~\cite{Pospelov:2005pr,Engel:2013lsa}. In general, operators with higher
dimension are suppressed by increasing inverse powers of $\Lambda_{\rm
  BSM}$ where $\Lambda_{\rm BSM}$ is the scale of new physics. The
couplings associated with these operators encode information about the
BSM model at the scale $\Lambda_{\rm BSM} \sim$~TeV and the renormalization
group evolution from $\Lambda_{\rm BSM}$ to the hadronic scale. 

The current goal of lattice QCD calculations is to examine \CPV\  operators with dimension six
and lower.  Of these, \CPV\  four
quark operators of dimension six have not been considered because they are sub-leading in
many BSM scenarios and because the lattice methodology to compute their
contribution to the neutron EDM has not yet been developed. 
Focus of the lattice community has been on the following four, which encode the
leading \CPV\  effects in a large class of BSM models:
\begin{eqnarray} 
{\cal L}_{\rm QCD}
    \mathbin{{\longrightarrow}} {\cal L}_{\rm
      QCD}^{\cancel{\text{CP}}} = {\cal L}_{\rm QCD} &+& i \Theta
    G_{\mu\nu} {\tilde G_{\mu\nu}} \ + \left. i \sum_q d_q^\gamma
    \overline{q} \sigma^{\mu\nu} {\tilde F_{\mu\nu}} q \right. +
    \left.  i \sum_q d_q^G \overline{q} \sigma^{\mu\nu} {\tilde
      G_{\mu\nu}} q \right.  \nonumber \\
  &+& {d_G} \ f^{abc} G_{\mu \nu}^a
    \tilde{G}^{\nu \beta,b} G_\beta^{\mu,c} 
\label{eq:Lcpv}
\end{eqnarray}
%  G_{\mu\nu} {\tilde G_{\mu\nu}} \ - \left. \frac{i}{2} \sum_q d_q^\gamma
%    \overline{q} \sigma^{\mu\nu} {\tilde F_{\mu\nu}} q \right. -
%    \left. \frac{i}{2} \sum_q d_q^G \overline{q} \sigma^{\mu\nu} {\tilde
%
%  
where the first term is the $\Theta-$interaction and the last is the
dimension six three-gluon Weinberg operator. The $\Theta$-term is a
part of the SM, but is usually neglected because the coupling $\Theta$
is constrained to be smaller than $10^{-10}$ by the current bound on
the nEDM and/or it is assumed that some form of a Peccei-Quinn
mechanism tunes $\Theta$ to zero~\cite{Peccei:1977hh}. Note that the
$\Theta$-term can be rotated into a pseudoscalar mass term $i
m_\ast({\Theta}) \ \overline{q} \gamma_5 q$ under a chiral
transformation~\cite{Crewther:1979pi}.  The two dimension five
operators are called the quark EDM (qEDM) and the quark chromo-EDM
(cEDM). The couplings $d_{u,d,s}^{\gamma}$ are the quark EDMs, the
$d_{u,d,s}^{g}$ are the quark chromo-EDMs, and $d_G$ is the strength
of the Weinberg operator.  They are generated directly by threshold
effects at the scale $\Lambda_{\rm BSM}$ or by mixing under
renormalization group evolution.  They parameterize the strength of
new CP violating interactions that a given BSM theory generates at the
hadronic scale.  

In an ideal world, the best way to calculate these matrix elements
would be to simulate a lattice theory with these \CPV\ interactions (Eq.~\eqref{eq:Lcpv}) added
to say the Wilson-clover fermion action.  This ideal approach does not
work because these interactions are complex and we do not yet know how
to efficiently simulate theories with a complex action.  The approach
that works for lattice QCD is to treat the small $d_q^{\gamma,g}$,
$\Theta$ and $d_G$ as perturbations and expand the theory about the
normal CP conserving action, such as the Wilson-clover action.  Then,
to lowest order in $\alpha_{em}$, the lattice calculation involves the product of
the electromagnetic current $J^{\rm EM}_\mu$ and each of these
operators: 
\begin{eqnarray}
  \left.\langle n \mid J^{\rm EM}_\mu \mid n \rangle\right|_{\not{\rm CP}}^{\Theta} &=&
       \left\langle n \left| J^{\rm EM}_\mu \ \int d^4 x\ \Theta  G^{\mu\nu}
       \tilde G^{\mu\nu} \right| n \right\rangle \,,
\label{eq:thetaEDM}
\\
  \left.\langle n \mid J^{\rm EM}_\mu \mid n \rangle\right|_{\not{\rm CP}}^{\rm qEDM} &=& 
    \epsilon_{\mu\nu\kappa\lambda} q^\nu
    \left\langle n     \left| \left( 
       d_u^\gamma \bar u \sigma^{\kappa\lambda} u + 
       d_d^\gamma \bar  d\sigma^{\kappa\lambda} d +  
       d_s^\gamma \bar  s\sigma^{\kappa\lambda} s\right) \right| n \right\rangle \,, 
\label{eq:quarkEDM}
\\
  \left.\langle n \mid J^{\rm EM}_\mu \mid n \rangle\right|_{\not{\rm CP}}^{\rm cEDM} &=& 
       \left\langle n \left| J^{\rm EM}_\mu \ \int d^4 x\ 
       \left( d_u^g \bar u \sigma_{\nu\kappa} u + 
              d_d^g \bar d \sigma_{\nu\kappa} d + 
              d_s^g \bar s \sigma_{\nu\kappa} s + \right) \tilde G^{\nu\kappa} \right| n \right\rangle \,,
\label{eq:chromoEDM}
\\
  \left.\langle n \mid J^{\rm EM}_\mu \mid n \rangle\right|_{\not{\rm CP}}^{G} &=&
       \left\langle n \left| J^{\rm EM}_\mu \ \int d^4 x\ {d_G} \, f^{abc} \, G_{\mu \nu}^a  \tilde{G}^{\nu \beta,b} G_\beta^{\mu,c}
       \right| n \right\rangle \,.
\label{eq:weinbergEDM}
\end{eqnarray}
The qEDM is an exception as the leading contribution to it arises from
the modification to $J^{\rm EM}_\mu$ as discussed in
Sec.~\ref{sec:qEDM}.  These matrix elements of $J^{\rm EM}_\mu$
between neutron states in the presence of CP violation are model
independent and provide the ``connection'' between the couplings and
the nEDM as exemplified in Eq.~\eqref{eq:dnqedm}.  Note that, since
each \CPV\ interaction contributes to the nEDM, the value of (bound
on) the nEDM provides a single constraint on the sum of all the
contributions. Nevertheless, lowering the bound on the nEDM will
provide increasingly tight, and amongst the most stringent constraint,
on possible BSM models.

%%%%%%%%%%%%%%%%%%%%%%%%%%%%%%%%%%%%%%%%%%%%%%%%%%%%%%%%%%%%%%%%%%%%%%%%%%%%
\section{Challenges to the calculation}
\label{sec:challenges}

The contributions of the $\Theta$, cEDM and Weinberg terms to the nEDM
are, in most lattice calculations, obtained from the $P$ and $T$
violating form factor $F_3$ that arises in the decomposition of the
matrix elements defined in
Eqs.~(\ref{eq:thetaEDM}),~(\ref{eq:chromoEDM}),
and~(\ref{eq:weinbergEDM}). For example, for the cEDM operator and 
ignoring the standard form factors $F_1$ and $F_2$ in the first term for 
brevity, one finds 
 \begin{eqnarray}
   \langle n | J^{\rm EM}_\mu | n \rangle_{\not{\rm CP}} = 
            \frac{F_3(q^2)}{2 M_n} \bar u_n\, q_\nu \sigma^{\mu\nu}\gamma_5\, u_n  \quad {\rm and} \quad
            d_n = \lim_{q^2\to0} \frac{F_3(q^2)}{2 M_n}
\label{eq:F3}
 \end{eqnarray}
In practice, in the extraction of $F_3$ from the matrix element, one
has to take into account mixing of $F_3$ with the $CP$ conserving form
factor $F_2$ due to a phase $e^{i \alpha_N \gamma_5}$ that
\CPV\ interaction generates in the neutron interpolating
operator. Note that there is a different $\alpha_N$ generated for each
of the \CPV\ interactions. This subtlety has been discussed and
resolved in Ref.~\cite{Abramczyk:2017oxr}.  To keep this review brief,
the reader is referred to the original papers cited for details. Here,
I assume that the reader is familiar with the correct phase
convention, the calculation of this phase $\alpha_N$ from the nucleon
2-point function, and its impact on the extraction of $F_3(Q^2)$.

There are two very important challenges to getting results at the physical point for the 
contribution of the $\Theta$, cEDM and Weinberg operators to the nEDM:
\begin{itemize}
\item
The first important challenge is the signal in $F_3$ is very small. In
fact, in most calculations, it is barely significant. The matrix
elements have to be calculate for non-zero momentum transfer,
excited-state contamination (ESC) removed, $F_3$ extracted, mixing with $F_2$ resolved,  and then 
extrapolated to $Q^2 = 0$.  Each of these steps is non-trivial.  In Secs.~\ref{sec:signal}
and~\ref{sec:cEDM}, I describe newly developed variance reduction
methods that reduce the errors by almost a factor of ten.
\item
Even after a signal has been demonstrated, defining the renormalized
cEDM and Weinberg operators and obtaining finite results in the
continuum limit is non-trivial. The reason is they mix with the same
and lower dimension operators under renormalization. The 1-loop
analysis of the cEDM operator~\cite{Bhattacharya:2016oqm} shows that
it mixes with the pseudoscalar $i\overline{q}\gamma_5 q$ and the
$\Theta$ operators.  Of these, the mixing with $i\overline{q}\gamma_5
q$ is quadratically divergent in all lattice formulations, i.e., both
chiral symmetry preserving and violating. As a result, the mixing
coefficients have to be determined very precisely
nonperturbatively. In fermion formulations such as Wilson-clover that
explicitly break chiral symmetry, there is an additional divergent
mixing with $i m \overline{q}\gamma_5 q$.  The mixing pattern of the
Weinberg operator is even more complicated.
\end{itemize}

%%%%%%%%%%%%%%%%%%%%%%%%%%%%%%%%%%%%%%%%%%%%%%%%%%%%%%%%%%%%%%%%%%%%%%%%%%%%%%%%%%%%%%%%%%%%
\begin{figure}[tpb] %1
\centerline{
\includegraphics[width=0.32\linewidth]{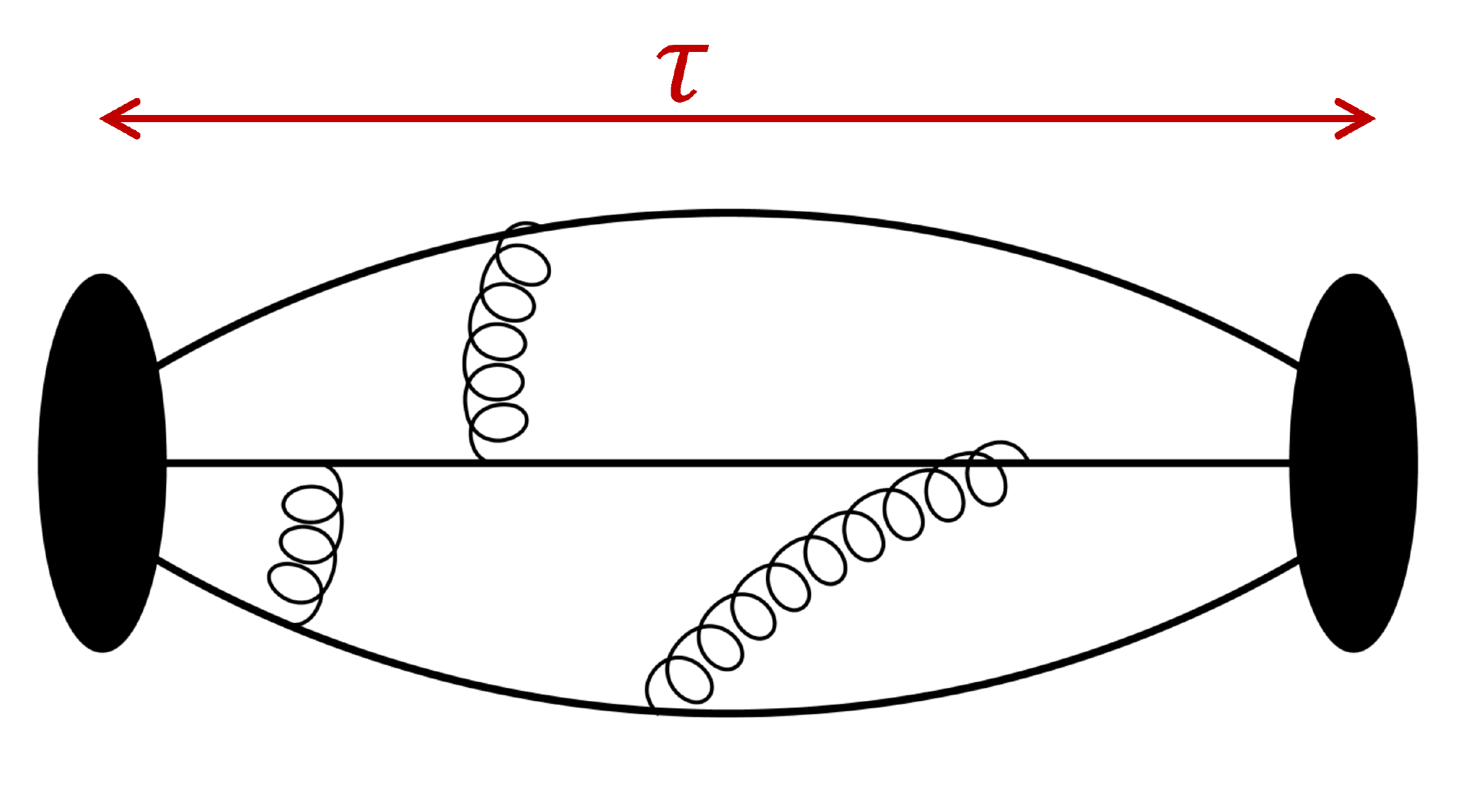} 
\includegraphics[width=0.32\linewidth]{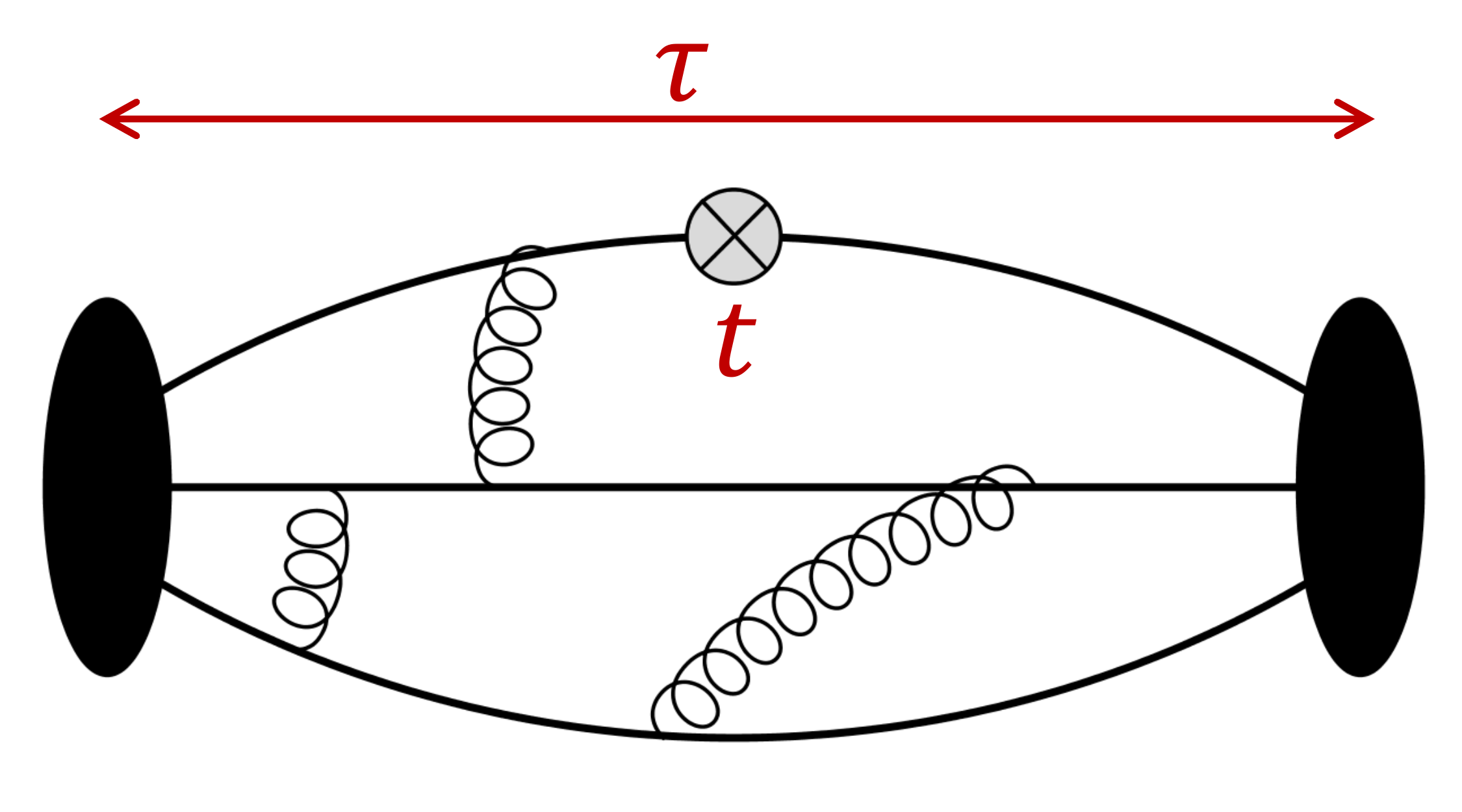}
\includegraphics[width=0.32\linewidth]{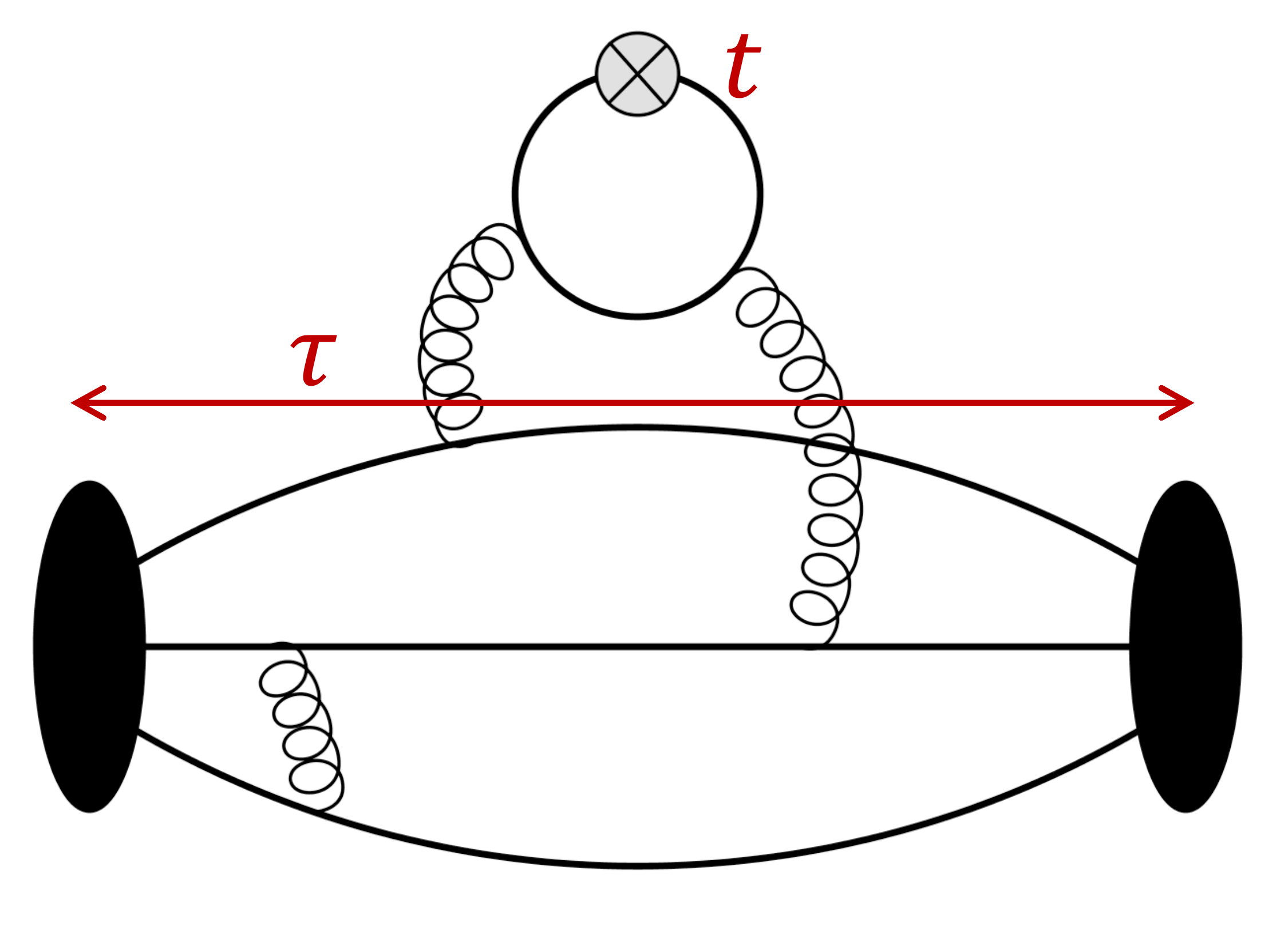}}
\caption{Illustration of the two- and three-point correlation
  functions calculated to calculate the flavor diagonal tensor charges 
  which give the contribution of the quark EDM operator to nEDM. 
  (Left) the nucleon two-point function. (Middle) the
  connected three-point function with source-sink separation $\tau$
  and tensor operator insertion on time slice $t$. (Right) the analogue
  disconnected three-point function that contributes to the flavor diagonal operators. }
\label{fig:con_disc}
\end{figure}
%
%
%%%%%%%%%%%%%%%%%%%%%%%%%%%%%%%%%%%%%%%%%%%%%%%%%%%%%%%%%%%%%%%%%%%%%%%%%%%%%%%%%%%%%%%%%%%%
\begin{figure}[tpb] %1
\centerline{
\includegraphics[width=0.32\linewidth]{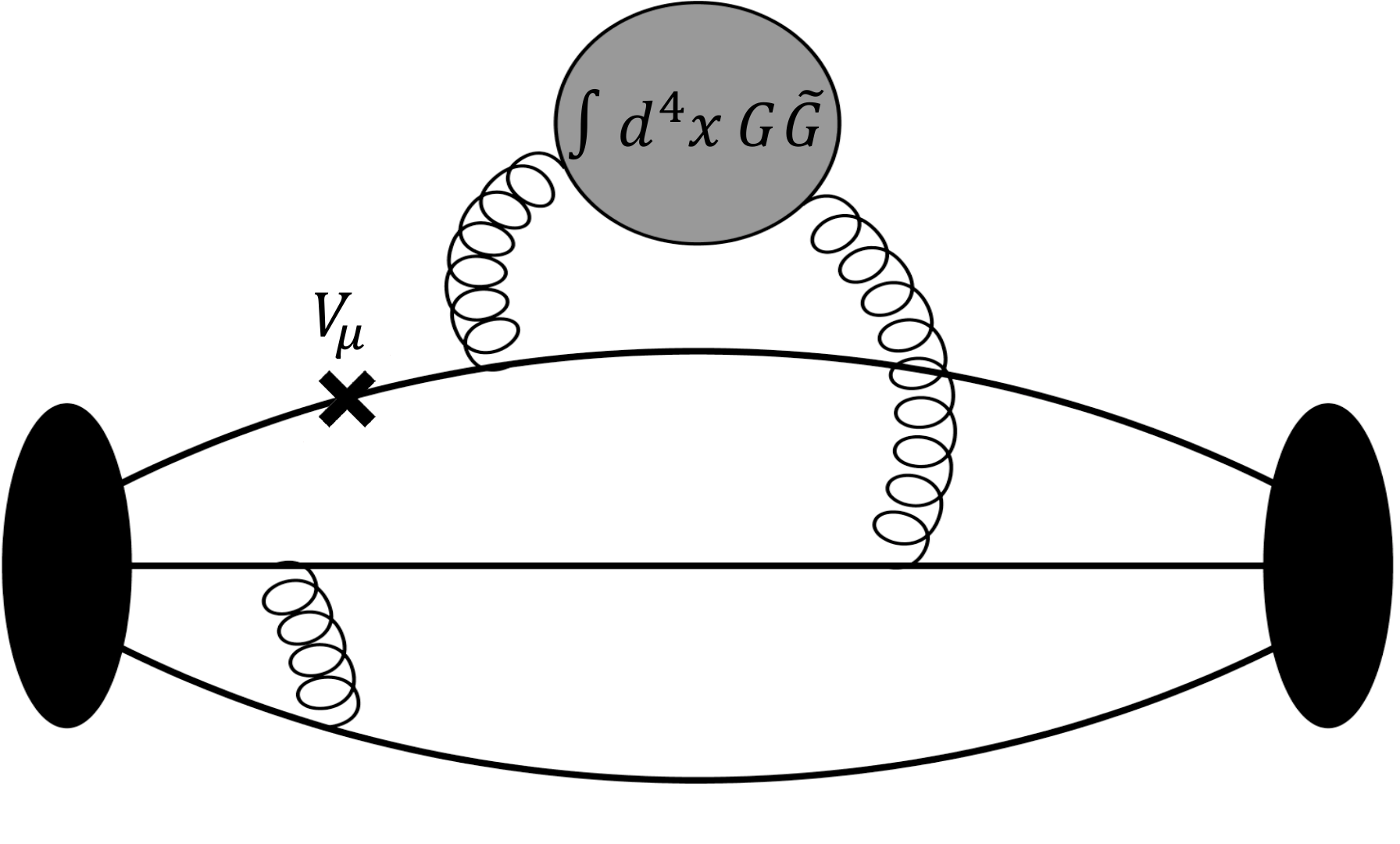} }
\caption{Illustration of the correlation of the $\Theta$ term with the
  nucleon 3-point function. The diagram for the Weinberg operator is
  the same with $ G \tilde G$ replaced by $ G G \tilde G$.
}
\label{fig:theta}
\end{figure}
%
%
%%%%%%%%%%%%%%%%%%%%%%%%%%%%%%%%%%%%%%%%%%%%%%%%%%%%%%%%%%%%%%%%%%%%%%%%%%%%%%%%%%%%%%%%%%%%

%%%%%%%%%%%%%%%%%%%%%%%%%%%%%%%%%%%%%%%%%%%%%%%%%%%%%%%%%%%%%%%%%
\section{Quark EDM operator and its matrix element}
\label{sec:qEDM}
%%%%%%%%%%%%%%%%%%%%%%%%%%%%%%%%%%%%%%%%%%%%%%%%%%%%%%%%%%%%%%%%%

In the presence of \CPV\ interactions, the electromagnetic current,
defined as $\delta {\cal L}/\delta A^\mu$ with $\cal L$ given in
Eq.~\eqref{eq:Lcpv}, gets an additional term
\begin{equation}
e \sum_q \overline{q} \gamma^{\mu} q \mathbin{{\longrightarrow}}
   e \sum_q \overline{q} \gamma^{\mu} q  + \epsilon^{\mu\nu\rho\sigma} p^\nu \sum_q d_q^\gamma \overline{q} \Sigma^{\rho\sigma} q \,, 
\label{eq:Vmu}
\end{equation}
which, at $\vec p = 0$ is the quark bilinear operator with tensor
structure. There is also a term analogous to Eq.~\eqref{eq:chromoEDM}
that is generated. It is, however, suppressed by $\alpha_{\rm em}$ and
therefore not considered. Thus, to leading order the contribution to neutron EDM
is given by
\begin{equation}
d_n = d_u^\gamma g_T^u + d_d^\gamma g_T^d + d_s^\gamma g_T^s, \cdots  \,,
\label{eq:dnqedm}
\end{equation}
where $g_T^{u,d,s}$ are the flavor diagonal tensor charges given by the matrix elements 
$\langle N | \overline{q} \sigma^{\mu\nu} q |N \rangle$ and
$d_{u,d,s}^\gamma$ are the corresponding BSM couplings.

The lattice methodology for the calculation of $g_T^{u,d,s}$ is robust
and reliable results at the physical pion mass $M_\pi=135$~MeV and in
the continuum limit have been presented in
Ref.~\cite{Gupta:2018lvp}. A brief description of the methodology of
lattice QCD calculations of nucleon 3-point functions for the
analogous axial operator is given in a companion paper in the
proceedings of this conference, PoS(Spin2018)018. It involves
calculating the three correlations functions illustrated in
Fig.~\ref{fig:con_disc}, with the qEDM (tensor)
operator inserted at time $t$ in the connected and disconnected 3-point diagrams.

Lattice calculations of nucleon charges have been reviewed in the FLAG
2019 report~\cite{Aoki:2019cca}. Data for the tensor charges exhibit 
small discretization or finite volume corrections and have
been stable over time. The current best
estimates, in the $\overline{MS}$ scheme at 2~GeV, are
\begin{equation}
g_T^u  = 0.784(28)(10); \quad g_T^d  = -0.204(11)(10); \quad g_T^s  = -0.0027(16) \,.
\label{eq:gT}
\end{equation}
Both the connected and disconnected contributions were obtained at the physical
point by fitting data at multiple values of $a$ and
$M_\pi$ and removing the leading continuum-chiral corrections. In short, 
flavor-diagonal tensor charges have been calculated with control
over all systematics~\cite{Gupta:2018lvp,Aoki:2019cca}.

Using these results, the authors of Ref.~\cite{Gupta:2018lvp} analyze
constraints on the split SUSY
model~\cite{ArkaniHamed:2004fb,Giudice:2004tc,ArkaniHamed:2004yi}. This
model is pertinent because in it qEDM is the dominant \CPV\  operator.
Using the experimental bounds $d_n \le 2.9 × 10^{−26}$ e
cm~\cite{Baker:2006ts} and $d_e \le 1.1 × 10^{−29}$ e
cm~\cite{Andreev:2018ayy}, gave the upper bound $d_n \le 4.1 ×
10^{−29}$ e cm for the split-SUSY model~\cite{Gupta:2018lvp}. More BSM theories can  be analyzed 
as results for other \CPV\ operators become available. 

\section{Calculation of the $\Theta$-term}
\label{sec:signal}

The $\Theta$-term breaks $P$ and $T$ invarinace and thus $CP$ by the
CPT theorem.  In the absence of a Peccei-Quinn mechanism, the
$\Theta$-term arises naturally in the SM. Operator mixing under
renormalization group flow of the cEDM and Weinberg operators between
$\Lambda_{\rm BSM}$ and the hadronic scale also generates it, i.e.,
the cEDM and Weinberg operators mix with the $\Theta$-term under
renormalization. Therefore, calculating its contribution to nEDM is
essential. Attributing the cause of a non-zero nEDM to an intrinsic
$\Theta$-term versus one generated from BSM interactions will be
challenging and will require knowing the matrix elements of all [leading]
\CPV\ operators.

A number of calculations of the contribution of the $\Theta$-term to
nEDM using the $F_3$ form factor method have been
done~\cite{Shintani:2005xg,Berruto:2005hg,Guo:2015tla,Alexandrou:2015spa,Shintani:2015vsx,Abramczyk:2017oxr,Syritsyn:2019vvt,Dragos:2019oxn}. Of
these, calculations done prior to Ref.~\cite{Abramczyk:2017oxr} have
poor [no] statistical signal and did not properly include the phase
induced by \CPV\ operators in the nucleon state as pointed out in
Ref.~\cite{Abramczyk:2017oxr}.

The calculation involves the correlation of the purely gluonic
operator $\int d^4xG^{\mu\nu} \tilde G^{\mu\nu}$ (topological charge)
with the nucleon 3-point function $ \langle N(0) J^{\rm EM}_\mu (t)
N(\tau) \rangle$ as shown in Fig.~\ref{fig:theta}.  Depending on the
lattice generation methodology, the value of the gluonic term, the
topological charge, can exhibit very long time auto-correlations. The
signal in the fermionic part is very good and typically $ \langle N(0)
J^{\rm EM}_4 (t)N(\tau) \rangle$ is used. The correlation between the
two is, therefore, the fermionic part weighted by the topological
charge. Consequently, if the topological charge is frozen during lattice 
generation, then the correlation and the projection on
to $F_3$ will have a poor/biased signal.

\begin{figure}[tpb] %1
\centerline{
\includegraphics[width=0.40\linewidth]{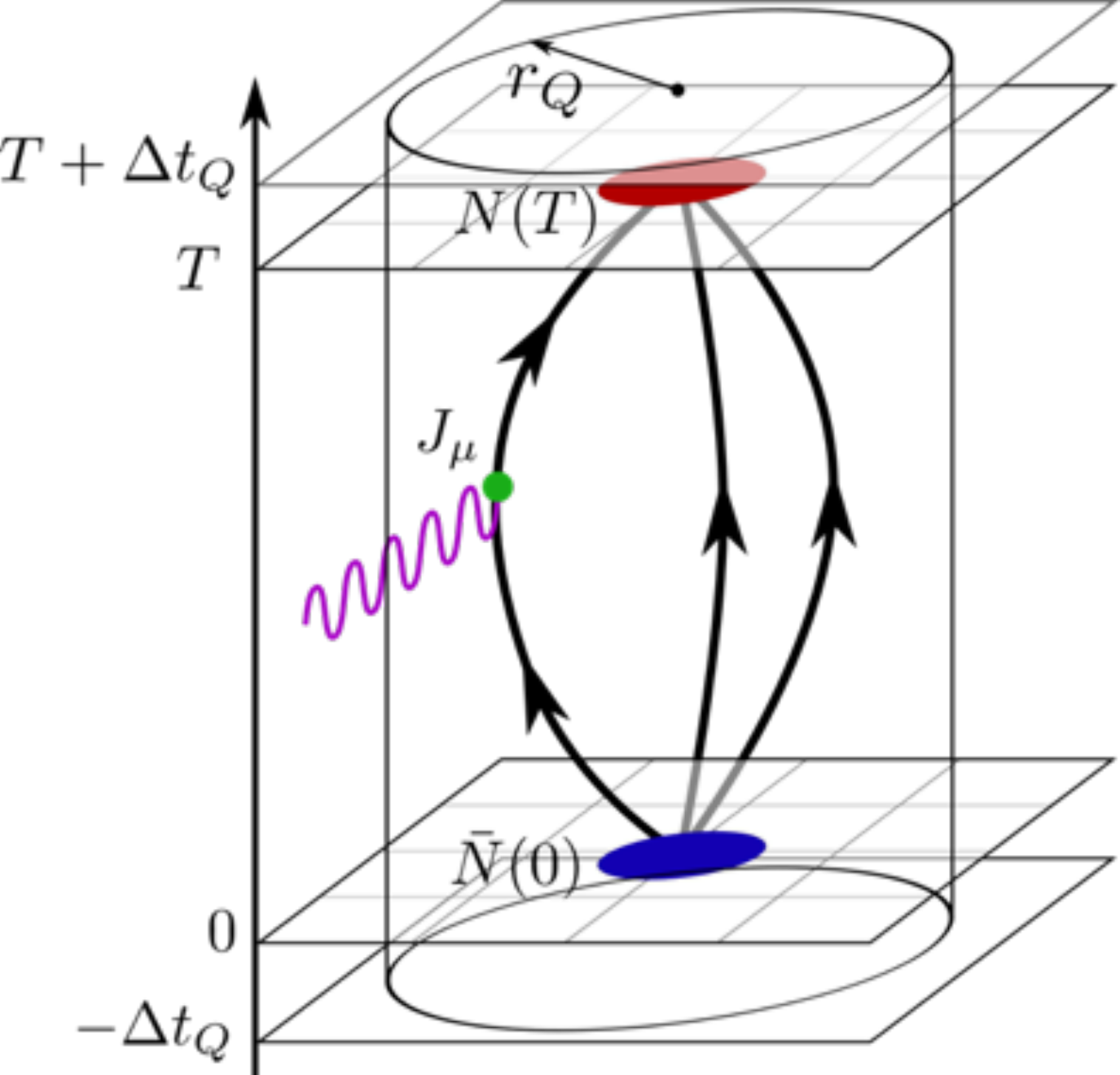} }
\caption{Using a 4-d cylinder about the nucleon 3-point correlator in
  which to sum $G^{\mu\nu} \tilde G^{\mu\nu}$ to reduce noise as
  explored in Ref.~\protect\cite{Syritsyn:2019vvt}.  The expectation
  is that points outside contribute only noise, however, the
  possibility that no bias has been introduced has not been established.  }
\label{fig:cylinder}
\end{figure}

Preliminary analysis with evidence of a signal has been presented in
Ref~\cite{Syritsyn:2019vvt}. This study used only one 2+1 flavor
domain wall ensemble with $a=0.1105$~fm and $M_\pi=340$~MeV;
calculated only the connected 3-point nucleon correlation function;
and did not remove the ESC. The topological charge was defined in the
gradient flow scheme to reduce noise in it.  Most of the effort has
been devoted to developing/testing the following variance redution
method to get a signal. In the correlation of $\int d^4x G^{\mu\nu}
\tilde G^{\mu\nu}$ with the nucleon 3-point function, the authors
propose to use a 4-d volume centered about the nucleon correlator over
which to sum $G^{\mu\nu} \tilde G^{\mu\nu}$ rather than the whole
lattice which gives the topological charge. This setup is illustrated
in Fig.~\ref{fig:cylinder}.  The motivation is that $G^{\mu\nu} \tilde
G^{\mu\nu}$ on points outside this volume contribute only noise.
While there is some evidence for variance reduction as a result, the
concomitant issue of introducing a possible bias has not been settled. One could
apply a variant of the standard bias correction
method~\cite{Bali:2009hu,Blum:2012uh} by calculating the correlation
with both the full and subvolume sum for a few of the nucleon 3-point
functions on each configuration. Such a bias correction method has not
yet been explored.

Furthermore, the authors contend that a reliable signal at the physical pion mass 
will need a new level of precision or alternate methods. 
Chiral perturbation theory predicts that the contribution of the $\Theta$-term to $d_n$ vanishes 
in the chiral limit as~\cite{Ottnad:2009jw,Mereghetti:2010kp,Mereghetti:2010tp}
\begin{equation}
d_N^\Theta =  a\, M_\pi^2 + b\, M_\pi^2 \ln M_\pi^2 + \cdots \,.
\end{equation}
Thus, as one tunes $M_\pi \to 135$~MeV in lattice simulations, the
precision of the calculation will have to be increased significantly
to keep the fractional error the same. For
example, on going from $340 \to 135$~MeV, the value of $d_N^\Theta
\propto M_\pi^2$ is expected to decrease by a factor of about six. Thus, the
statistics will have to be increased by $O(100)$ to keep the fractional
error the same~\cite{Syritsyn:2019vvt}.

Results from six ensembles with $M_\pi \ge 410$~MeV have just been
published in Ref.~\cite{Dragos:2019oxn}.  The topological charge
density is again calculated in the gradient flow scheme. The central
values are all negative, lie between $-0.0021$ and $-0.0070$, but five
of them differ by less than 2$\sigma$ from zero. There is no evidence
of a $d_N^\Theta \propto M_\pi^2$ behavior at these heavy pion masses,
and a less than convincing chiral-continuum fit gave $d_n =
-0.00186(59) \ \Theta$ e fm. Clearly, far more precise calculations
near the physical pion mass are needed before a result can be
presented with confidence.

My bottom line conclusion on the contribution of the $\Theta$-term is
that, while, the methodology for the calculation of $F_3$ from the
$\Theta$-term is now established and there are no show stopping issues
of renormalization, more work needs to be done to demonstrate a
$5\sigma$ signal on $M_\pi \gtrsim 350$~MeV ensembles, and $O(100)$
more to obtain results at $M_\pi = 135$~MeV and in the $a \to 0$
limit. \looseness-1

\section{Enhancing the signal in $F_3$ from the cEDM operator}
\label{sec:cEDM}

Calculations of the cEDM operator need to address both the signal and
the renormalization, including the divergent mixing with the
pseudoscalar operator $i\overline{q}\gamma_5 q$. The two calculations
reported in Refs.~\cite{Bhattacharya:2018qat,Syritsyn:2019vvt} are
both based on extracting $d_n$ from $F_3$. So far, the focus in these
works has been on getting a signal, and that too in only the connected part.  No results have been reported for
the mixing coefficients calculated non-perturbatively in either of the
two approaches. \looseness-1

The work described in Ref.~\cite{Bhattacharya:2018qat} uses the
Schwinger source method to include the \CPV\ interaction in the Dirac
action.  The authors show that the phase $\alpha_N$ associated with
the neutron ground state and generated by the \CPV\ interaction can be
extracted reliably from the 2-point function. To obtain a signal in
$F_3$, which $a\ priori$ was poor, they developed a general variance
reduction method. It exploits correlations between an observable $O$
and a number of other quantities $R_i$ with $\langle R_i \rangle =
0$. For the connected part of cEDM, the variance of the combination $O
+ \xi_i R_i$ was shown to be reduced by almost a factor of ten by a suitable 
choice of $R_i$ and optimizing the parameters $\xi_i$.

The extraction of the three form factors, $F_1$, $F_2$ and $F_3$, from
the 8 matrix elements (real and imaginary parts of the four components
of the vector current) forms an over complete set. The authors find
that the results for the unrenormalized connected contributions to
$F_3$ from different combinations give different estimates. These
differences point to possible large excited state and/or
discretization effects that need to be resolved.  The real hurdle in
these calculations, in schemes such as RI-sMOM, is controlling the
divergent mixing with the lower dimension pseudoscalar operator.  Work
has, therefore, been initiated to do the calculation in the gradient
flow scheme.

Results for the unrenormalized connected contributions to both the
cEDM and pseudoscalar operator it mixes with using the direct 4-point
method have been presented in Ref.~\cite{Syritsyn:2019vvt}.  Having
demonstrated a signal in $F_3$ for both the cEDM and pseudoscalar
operator, they are now investigating the position-space
renormalization scheme to control the mixing problem, and on including
the disconnected contributions.

\section{The Weinberg operator}
\label{sec:Weinberg}

\begin{figure}[tpb] %4
\centerline{
\includegraphics[width=0.40\linewidth]{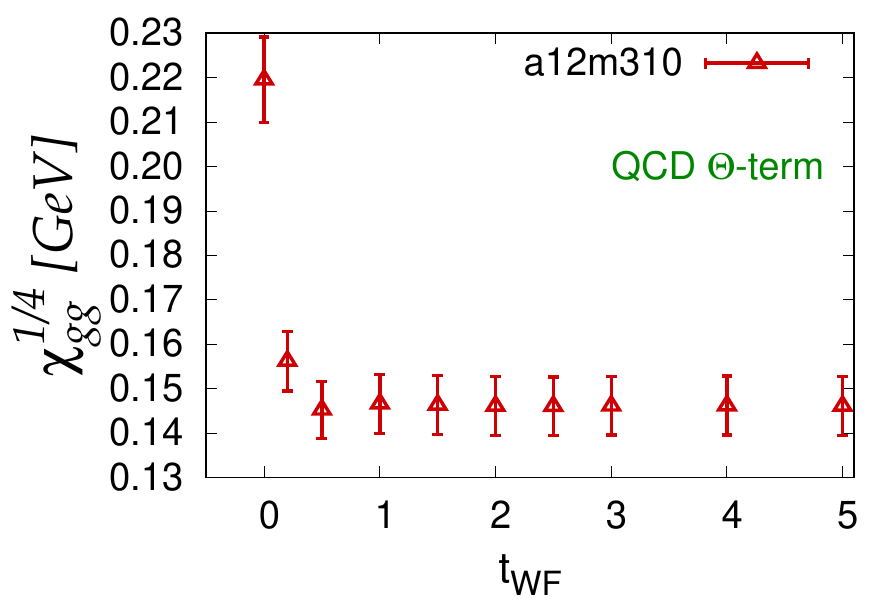} 
\includegraphics[width=0.40\linewidth]{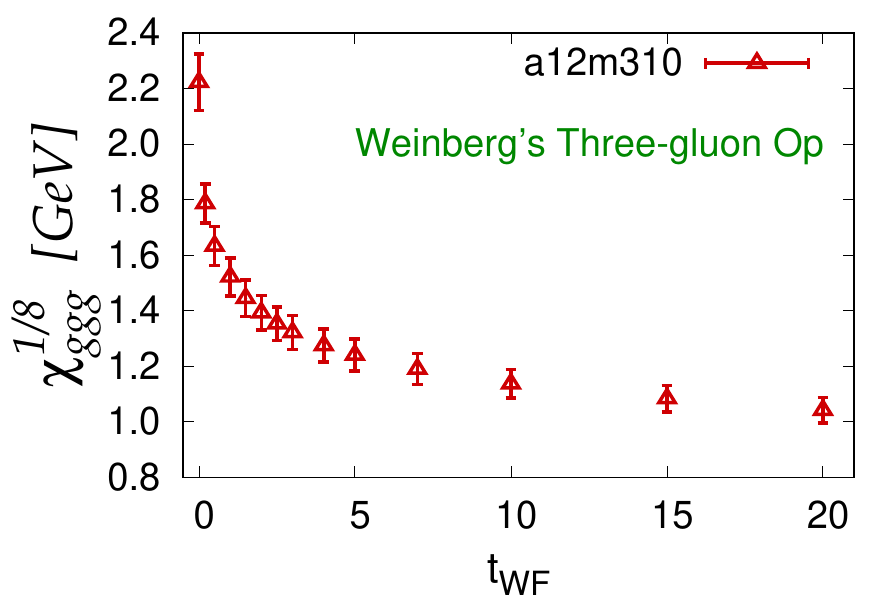}}
\caption{Susceptibility of topological charge (left) and 
Weinberg operator (right) versus the flow time $t_{\rm WF}$. 
}
\label{fig:Weinbergsus}
\end{figure}

The lattice calculation of the Weinberg operator is similar to that
of the $\Theta$-term except for the additional serious complication of
a divergent mixing with the $\Theta$-term under renormalization.  The
method currently being explored that would control this mixing on the
lattice is gradient flow~\cite{Rizik:2018lrz,Bhattacharya:2018qat}.
Fig.~\ref{fig:Weinbergsus} shows a comparison of the
susceptibilities of the topological charge and the Weinberg operator
as a function of the flow time. The data for the topological
susceptibility quickly flattens out as expected while that for the
scale-dependent Weinberg susceptibility continues to evolve. The next
step is to calculate the 4-point correlation function as a function of
the flow time. Simultaneously, our extended collaboration at LANL is
working to relate the gradient flow scheme to the continuum
$\overline{MS}$ scheme. With this matching in hand, we will need to
demonstrate that there exists a window in flow time in which results
in the continuum are independent of the flow time. In short, a number
of developments have to pan out before physical results can be
obtained.

\begin{figure}[tpb] %5
\centerline{
\includegraphics[height=0.30\linewidth]{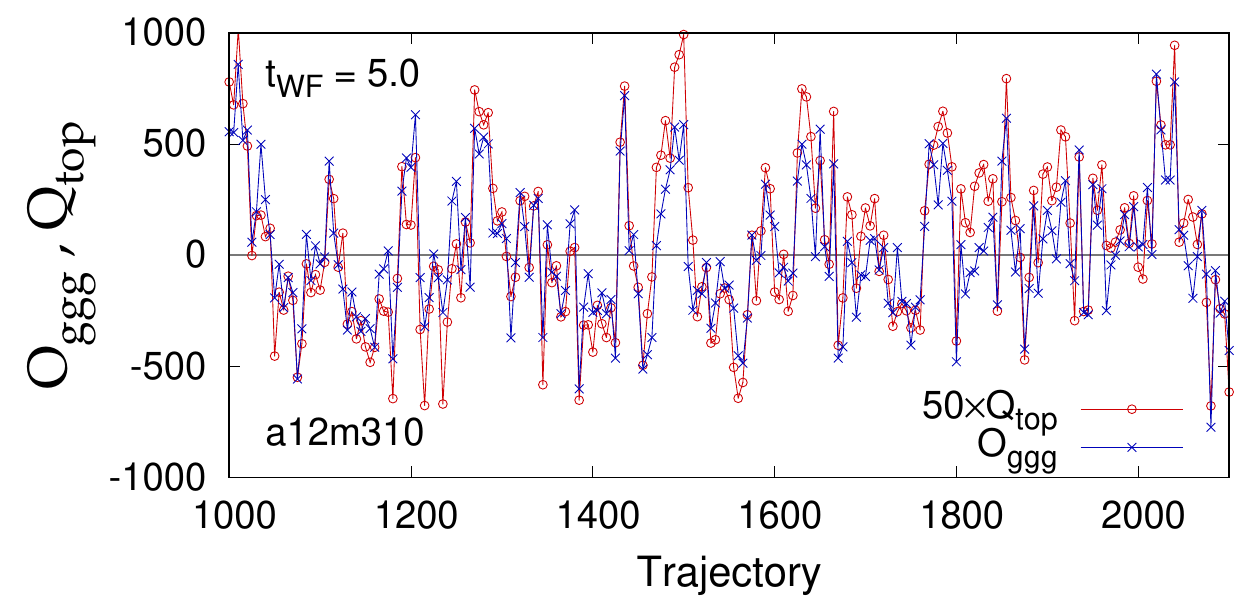} 
}
\caption{The data for the topological charge (red) and the Weinberg
  operator (blue), at flow time $t_{\rm WF} = 5$, plotted as a
  function of the Monte Carlo configuration number. The data show very
  strong correlations between the two for sufficiently large flow
  time, in this case for $t_{\rm WF} \gtrsim 0.5$. }
\label{fig:Weinberg}
\end{figure}

Fig.~\ref{fig:Weinberg} shows data for the Weinberg operator and the
topological charge versus Monte Carlo configuration number on
$a12m310$ ensemble at flow time $t_{\rm WF} = 5$.  The two are highly
correlated. So, any technique for improving
the signal in the $\Theta$-term will also be applicable to the
Weinberg operator. It also highlights a large divergent mixing with the
$\Theta$-term that needs to be addressed. \looseness-1

%%%%%%%%%%%%%%%%%%%%%%%%%%%%%%%%%%%%%%%%%%%%%%%%%%%%%%%%%%%%%%%%%%%%%%%%%%%%%%%%%%%%%%%%%%%%
%
%
%%%%%%%%%%%%%%%%%%%%%%%%%%%%%%%%%%%%%%%%%%%%%%%%%%%%%%%%%%%%%%%%%%%%%%%%%%%%%%%%%%%%%%%%%%%%
%
%
%%%%%%%%%%%%%%%%%%%%%%%%%%%%%%%%%%%%%%%%%%%%%%%%%%%%%%%%%%%%%%%%%%%%%%%%%%%%%%%%%%%%%%%%%%%%

\section{Conclusions}
\label{sec:conclusions}

The prospect of reducing the upper bound on the nEDM from current and
future experiments, and possibly finding a value is exciting. It will
signal $T$ (and $CP$ assuming CPT) violation larger than in the
standard model, and will put stringent constraints on BSM theories
provided the matrix elements of \CPV\ operators can be calculated
reliably. A number of groups are using large scale simulations of
lattice QCD to calculate the matrix elements of the four operators 
reviewed, and their contributions to the nEDM. These calculations are extremely hard except
for the quark EDM operator. Clearly new ideas and algorithms are
needed! My summary of the status and prospects for obtaining results of
dimension six and less operators, in order of the most likely within
the next five years, is the following:\looseness-1
\begin{itemize}
\item
The quark EDM operator. The leading contribution of this operator has
been calculated and results with $O(5\%)$ errors, after extrapolation
to the continuum limit and at the physical pion mass, have been
obtained as discussed in Sec.~\ref{sec:qEDM}.
\item
The $\Theta$-term: Calculations of its contribution to nEDM have the
longest history.  Evidence of a signal is growing, and I expect
validated results with $O(50\%)$ uncertainty to be available over the next 
five years.
\item
The chromo EDM operator: Methods to get a $5\sigma$ signal are being
developed and tested. Results including both the connected and
disconnected contributions at multiple values of the lattice spacing
require $O(100)$ or more in computing resources. Thereafter, tests and control
of the divergent mixing with the pseudoscalar operator can begin.
\item
The Weinberg operator: The mixing problem under renormalization is
severe. Working in the gradient flow scheme is the approach of choice
being investigated~\cite{Rizik:2018lrz,Bhattacharya:2018qat}. First
tests of the methodology and the numerical signal are being performed.
For both the Weinberg and chromo EDM operators, estimates with $O(1) $
uncertainty are possible in five years if the gradient flow method
works.
\item
4-fermion operators: So far there is no published work on a
renormalization framework for these operators that will be suitable
for lattice calculations nor have exploratory lattice calculations
begun. We are unlikely to see significant calculations over the next
five years, if for no other reason than due to the limited access to
computer time. Most groups will likely channel focus and resources
to the previous three operators first.
\end{itemize}

\acknowledgments
I thank Andreas Wirzba, Paolo Lenisa and the organizers of Spin 2018
for inviting me to give this review and their hospitality.  I thank
Tanmoy Bhattacharya, Vincenzo Cirigliano, Andrew Kobach, Emanuelle
Mereghetti and Boram Yoon, for a fun and fruitful collaboration in
which we have seen the statistical signal come and go, and the methods
evolve in response.  We thank the MILC collaboration for sharing the
$2+1+1$-flavor HISQ ensembles generated by them. We gratefully
acknowledge the computing facilities at, and resources provided by,
NERSC, OLCF at Oak Ridge, USQCD and LANL Institutional Computing.

%%%%%%%%%%%%%%%%%%%%%%%%%%%%%%%%%%%%%%%%%%%%%%%%%%%%%%%%%%%%%%%%%%%%%%%
\bibliographystyle{JHEP}
\bibliography{ref} %%% ref.bib file

\providecommand{\href}[2]{#2}\begingroup\raggedright\begin{thebibliography}{10}

\bibitem{Bennett:2003bz}
{\scshape WMAP Collaboration} collaboration, C.~Bennett et~al., \emph{{First
  year Wilkinson Microwave Anisotropy Probe (WMAP) observations: Preliminary
  maps and basic results}},
  \href{https://doi.org/10.1086/377253}{\emph{Astrophys.J.Suppl.} {\bfseries
  148} (2003) 1} [\href{https://arxiv.org/abs/astro-ph/0302207}{{\ttfamily
  astro-ph/0302207}}].

\bibitem{Kolb:1990vq}
E.~W. Kolb and M.~S. Turner, \emph{{The Early Universe}}, {\emph{Front.Phys.}
  {\bfseries 69} (1990) 1}.

\bibitem{Coppi:2004za}
P.~Coppi, \emph{{How Do We know Antimater is Absent?}}, {\emph{eConf}
  {\bfseries C040802} (2004) L017}.

\bibitem{Sakharov:1967dj}
A.~Sakharov, \emph{{Violation of CP Invariance, c Asymmetry, and Baryon
  Asymmetry of the Universe}},
  \href{https://doi.org/10.1070/PU1991v034n05ABEH002497}{\emph{Pisma
  Zh.Eksp.Teor.Fiz.} {\bfseries 5} (1967) 32}.

\bibitem{Kobayashi:1973fv}
M.~Kobayashi and T.~Maskawa, \emph{{CP Violation in the Renormalizable Theory
  of Weak Interaction}},
  \href{https://doi.org/10.1143/PTP.49.652}{\emph{Prog.Theor.Phys.} {\bfseries
  49} (1973) 652}.

\bibitem{Nunokawa:2007qh}
H.~Nunokawa, S.~J. Parke and J.~W. Valle, \emph{{CP Violation and Neutrino
  Oscillations}},
  \href{https://doi.org/10.1016/j.ppnp.2007.10.001}{\emph{Prog.Part.Nucl.Phys.}
  {\bfseries 60} (2008) 338} [\href{https://arxiv.org/abs/0710.0554}{{\ttfamily
  0710.0554}}].

\bibitem{Dolgov:1991fr}
A.~Dolgov, \emph{{NonGUT baryogenesis}},
  \href{https://doi.org/10.1016/0370-1573(92)90107-B}{\emph{Phys.Rept.}
  {\bfseries 222} (1992) 309}.

\bibitem{Pospelov:2005pr}
M.~Pospelov and A.~Ritz, \emph{{Electric dipole moments as probes of new
  physics}}, \href{https://doi.org/10.1016/j.aop.2005.04.002}{\emph{Annals
  Phys.} {\bfseries 318} (2005) 119}
  [\href{https://arxiv.org/abs/hep-ph/0504231}{{\ttfamily hep-ph/0504231}}].

\bibitem{Engel:2013lsa}
J.~Engel, M.~J. Ramsey-Musolf and U.~van Kolck, \emph{{Electric Dipole Moments
  of Nucleons, Nuclei, and Atoms: The Standard Model and Beyond}},
  \href{https://doi.org/10.1016/j.ppnp.2013.03.003}{\emph{Prog. Part. Nucl.
  Phys.} {\bfseries 71} (2013) 21}
  [\href{https://arxiv.org/abs/1303.2371}{{\ttfamily 1303.2371}}].

\bibitem{Peccei:1977hh}
R.~D. Peccei and H.~R. Quinn, \emph{{CP Conservation in the Presence of
  Instantons}}, \href{https://doi.org/10.1103/PhysRevLett.38.1440}{\emph{Phys.
  Rev. Lett.} {\bfseries 38} (1977) 1440}.

\bibitem{Crewther:1979pi}
R.~J. Crewther, P.~Di~Vecchia, G.~Veneziano and E.~Witten, \emph{{Chiral
  Estimate of the Electric Dipole Moment of the Neutron in Quantum
  Chromodynamics}}, \href{https://doi.org/10.1016/0370-2693(80)91025-4,
  10.1016/0370-2693(79)90128-X}{\emph{Phys. Lett.} {\bfseries 88B} (1979) 123}.

\bibitem{Abramczyk:2017oxr}
M.~Abramczyk, S.~Aoki, T.~Blum, T.~Izubuchi, H.~Ohki and S.~Syritsyn,
  \emph{{Lattice calculation of electric dipole moments and form factors of the
  nucleon}}, \href{https://doi.org/10.1103/PhysRevD.96.014501}{\emph{Phys.
  Rev.} {\bfseries D96} (2017) 014501}
  [\href{https://arxiv.org/abs/1701.07792}{{\ttfamily 1701.07792}}].

\bibitem{Bhattacharya:2016oqm}
T.~Bhattacharya, V.~Cirigliano, R.~Gupta, E.~Mereghetti and B.~Yoon,
  \emph{{Neutron Electric Dipole Moment from quark Chromoelectric Dipole
  Moment}}, \href{https://doi.org/10.22323/1.251.0238}{\emph{PoS} {\bfseries
  LATTICE2015} (2016) 238} [\href{https://arxiv.org/abs/1601.02264}{{\ttfamily
  1601.02264}}].

\bibitem{Gupta:2018lvp}
R.~Gupta, B.~Yoon, T.~Bhattacharya, V.~Cirigliano, Y.-C. Jang and H.-W. Lin,
  \emph{{Flavor diagonal tensor charges of the nucleon from (2+1+1)-flavor
  lattice QCD}}, \href{https://doi.org/10.1103/PhysRevD.98.091501}{\emph{Phys.
  Rev.} {\bfseries D98} (2018) 091501}
  [\href{https://arxiv.org/abs/1808.07597}{{\ttfamily 1808.07597}}].

\bibitem{Aoki:2019cca}
{\scshape Flavour Lattice Averaging Group} collaboration, S.~Aoki et~al.,
  \emph{{FLAG Review 2019}},
  \href{https://arxiv.org/abs/1902.08191}{{\ttfamily 1902.08191}}.

\bibitem{ArkaniHamed:2004fb}
N.~Arkani-Hamed and S.~Dimopoulos, \emph{{Supersymmetric unification without
  low energy supersymmetry and signatures for fine-tuning at the LHC}},
  \href{https://doi.org/10.1088/1126-6708/2005/06/073}{\emph{JHEP} {\bfseries
  0506} (2005) 073} [\href{https://arxiv.org/abs/hep-th/0405159}{{\ttfamily
  hep-th/0405159}}].

\bibitem{Giudice:2004tc}
G.~Giudice and A.~Romanino, \emph{{Split supersymmetry}},
  \href{https://doi.org/10.1016/j.nuclphysb.2004.11.048}{\emph{Nucl.Phys.}
  {\bfseries B699} (2004) 65}
  [\href{https://arxiv.org/abs/hep-ph/0406088}{{\ttfamily hep-ph/0406088}}].

\bibitem{ArkaniHamed:2004yi}
N.~Arkani-Hamed, S.~Dimopoulos, G.~Giudice and A.~Romanino, \emph{{Aspects of
  split supersymmetry}},
  \href{https://doi.org/10.1016/j.nuclphysb.2004.12.026}{\emph{Nucl.Phys.}
  {\bfseries B709} (2005) 3}
  [\href{https://arxiv.org/abs/hep-ph/0409232}{{\ttfamily hep-ph/0409232}}].

\bibitem{Baker:2006ts}
C.~Baker, D.~Doyle, P.~Geltenbort, K.~Green, M.~van~der Grinten et~al.,
  \emph{{An Improved experimental limit on the electric dipole moment of the
  neutron}},
  \href{https://doi.org/10.1103/PhysRevLett.97.131801}{\emph{Phys.Rev.Lett.}
  {\bfseries 97} (2006) 131801}
  [\href{https://arxiv.org/abs/hep-ex/0602020}{{\ttfamily hep-ex/0602020}}].

\bibitem{Andreev:2018ayy}
{\scshape ACME} collaboration, V.~Andreev et~al., \emph{{Improved limit on the
  electric dipole moment of the electron}},
  \href{https://doi.org/10.1038/s41586-018-0599-8}{\emph{Nature} {\bfseries
  562} (2018) 355}.

\bibitem{Shintani:2005xg}
E.~Shintani, S.~Aoki, N.~Ishizuka, K.~Kanaya, Y.~Kikukawa, Y.~Kuramashi et~al.,
  \emph{{Neutron electric dipole moment from lattice QCD}},
  \href{https://doi.org/10.1103/PhysRevD.72.014504}{\emph{Phys. Rev.}
  {\bfseries D72} (2005) 014504}
  [\href{https://arxiv.org/abs/hep-lat/0505022}{{\ttfamily hep-lat/0505022}}].

\bibitem{Berruto:2005hg}
F.~Berruto, T.~Blum, K.~Orginos and A.~Soni, \emph{{Calculation of the neutron
  electric dipole moment with two dynamical flavors of domain wall fermions}},
  \href{https://doi.org/10.1103/PhysRevD.73.054509}{\emph{Phys. Rev.}
  {\bfseries D73} (2006) 054509}
  [\href{https://arxiv.org/abs/hep-lat/0512004}{{\ttfamily hep-lat/0512004}}].

\bibitem{Guo:2015tla}
F.~K. Guo, R.~Horsley, U.~G. Meissner, Y.~Nakamura, H.~Perlt, P.~E.~L. Rakow
  et~al., \emph{{The electric dipole moment of the neutron from 2+1 flavor
  lattice QCD}},
  \href{https://doi.org/10.1103/PhysRevLett.115.062001}{\emph{Phys. Rev. Lett.}
  {\bfseries 115} (2015) 062001}
  [\href{https://arxiv.org/abs/1502.02295}{{\ttfamily 1502.02295}}].

\bibitem{Alexandrou:2015spa}
C.~Alexandrou, A.~Athenodorou, M.~Constantinou, K.~Hadjiyiannakou, K.~Jansen,
  G.~Koutsou et~al., \emph{{Neutron electric dipole moment using $N_{f} =2+1+1$
  twisted mass fermions}},
  \href{https://doi.org/10.1103/PhysRevD.93.074503}{\emph{Phys. Rev.}
  {\bfseries D93} (2016) 074503}
  [\href{https://arxiv.org/abs/1510.05823}{{\ttfamily 1510.05823}}].

\bibitem{Shintani:2015vsx}
E.~Shintani, T.~Blum, T.~Izubuchi and A.~Soni, \emph{{Neutron and proton
  electric dipole moments from $N_f=2+1$ domain-wall fermion lattice QCD}},
  \href{https://doi.org/10.1103/PhysRevD.93.094503}{\emph{Phys. Rev.}
  {\bfseries D93} (2016) 094503}
  [\href{https://arxiv.org/abs/1512.00566}{{\ttfamily 1512.00566}}].

\bibitem{Syritsyn:2019vvt}
S.~Syritsyn, T.~Izubuchi and H.~Ohki, \emph{{Calculation of Nucleon Electric
  Dipole Moments Induced by Quark Chromo-Electric Dipole Moments and the QCD
  $\theta$-term}},  in \emph{{13th Conference on Quark Confinement and the
  Hadron Spectrum (Confinement XIII) Maynooth, Ireland, July 31-August 6,
  2018}}, 2019, \href{https://arxiv.org/abs/1901.05455}{{\ttfamily
  1901.05455}}.

\bibitem{Dragos:2019oxn}
J.~Dragos, T.~Luu, A.~Shindler, J.~de~Vries and A.~Yousif, \emph{{Confirming
  the Existence of the strong CP Problem in Lattice QCD with the Gradient
  Flow}},  \href{https://arxiv.org/abs/1902.03254}{{\ttfamily 1902.03254}}.

\bibitem{Bali:2009hu}
G.~S. Bali, S.~Collins and A.~Schafer, \emph{{Effective noise reduction
  techniques for disconnected loops in Lattice QCD}},
  \href{https://doi.org/10.1016/j.cpc.2010.05.008}{\emph{Comput.Phys.Commun.}
  {\bfseries 181} (2010) 1570}
  [\href{https://arxiv.org/abs/0910.3970}{{\ttfamily 0910.3970}}].

\bibitem{Blum:2012uh}
T.~Blum, T.~Izubuchi and E.~Shintani, \emph{{New class of variance-reduction
  techniques using lattice symmetries}},
  \href{https://doi.org/10.1103/PhysRevD.88.094503}{\emph{Phys.Rev.} {\bfseries
  D88} (2013) 094503} [\href{https://arxiv.org/abs/1208.4349}{{\ttfamily
  1208.4349}}].

\bibitem{Ottnad:2009jw}
K.~Ottnad, B.~Kubis, U.~G. Meissner and F.~K. Guo, \emph{{New insights into the
  neutron electric dipole moment}},
  \href{https://doi.org/10.1016/j.physletb.2010.03.005}{\emph{Phys. Lett.}
  {\bfseries B687} (2010) 42}
  [\href{https://arxiv.org/abs/0911.3981}{{\ttfamily 0911.3981}}].

\bibitem{Mereghetti:2010kp}
E.~Mereghetti, J.~de~Vries, W.~H. Hockings, C.~M. Maekawa and U.~van Kolck,
  \emph{{The Electric Dipole Form Factor of the Nucleon in Chiral Perturbation
  Theory to Sub-leading Order}},
  \href{https://doi.org/10.1016/j.physletb.2010.12.018}{\emph{Phys. Lett.}
  {\bfseries B696} (2011) 97}
  [\href{https://arxiv.org/abs/1010.4078}{{\ttfamily 1010.4078}}].

\bibitem{Mereghetti:2010tp}
E.~Mereghetti, W.~H. Hockings and U.~van Kolck, \emph{{The Effective Chiral
  Lagrangian From the Theta Term}},
  \href{https://doi.org/10.1016/j.aop.2010.03.005}{\emph{Annals Phys.}
  {\bfseries 325} (2010) 2363}
  [\href{https://arxiv.org/abs/1002.2391}{{\ttfamily 1002.2391}}].

\bibitem{Bhattacharya:2018qat}
T.~Bhattacharya, B.~Yoon, R.~Gupta and V.~Cirigliano, \emph{{Neutron Electric
  Dipole Moment from Beyond the Standard Model}},  2018,
  \href{https://arxiv.org/abs/1812.06233}{{\ttfamily 1812.06233}}.

\bibitem{Rizik:2018lrz}
M.~Rizik, C.~Monahan and A.~Shindler, \emph{{Renormalization of CP-Violating
  Pure Gauge Operators in Perturbative QCD Using the Gradient Flow}},  in
  \emph{{36th International Symposium on Lattice Field Theory (Lattice 2018)
  East Lansing, MI, United States, July 22-28, 2018}}, 2018,
  \href{https://arxiv.org/abs/1810.05637}{{\ttfamily 1810.05637}}.

\end{thebibliography}\endgroup
%%%%%%%%%%%%%%%%%%%%%%%%%%%%%%%%%%%%%%%%%%%%%%%%%%%%%%%%%%%%%%%%%%%%%%%

%%\begin{thebibliography}{99}
%%\bibitem{...}
%% \end{thebibliography}

\end{document}